\documentstyle[a4wide,epsf,11pt,titlepage]{article}

\newcounter{nref}
\newcommand{\bbib}{%
  \renewcommand{\refname}{\large\bf References}%
  \begin{thebibliography}{99}%
  \setcounter{enumiv}{\thenref}}
\newcommand{\ebib}{%
  \end{thebibliography}%
  \setcounter{nref}{\arabic{enumiv}}}
\newcommand{\head}[3]{%
  \setcounter{nref}{0}%
  \thispagestyle{empty}%
  \section*{\LARGE\bf #1}%
  \stepcounter{section}%
  \addcontentsline{toc}{section}{#1}%
  \large\itshape%
  #2\\\vspace{0.1pt}\\%
  #3%
  \normalsize\upshape%
  \bigskip}

\begin{document}


\head{Numerical Simulations of Type Ia Supernova\\Explosions}
     {F.K.\ R{\"o}pke$^1$, W.\ Hillebrandt$^1$, M.\ Gieseler$^1$, M.\
     Reinecke$^1$, and C.\ Travaglio$^2$}
     {$^1$ Max-Planck-Institut f\"ur Astrophysik,
     Karl-Schwarzschild-Str.~1, 85741 Garching,\\Germany\\
      $^2$ INAF -- Osservatorio Astronomico di Torino,
              Strada dell'Osservatorio 20,\\I-10025 Pino Torinese,
              Torino, Italy}

\subsection{Introduction}
Recent numerical simulations of Type Ia Supernova (SN Ia)
explosions \cite{roepke.1, roepke.2} have successfully
modeled the Chandrasekhar-mass deflagration scenario (for a
review see \cite{roepke.3} or J.~Niemeyer's contribution to
these proceedings) in three spatial
dimensions. In this SN Ia model a carbon/oxygen white dwarf (WD) star
accretes matter from a binary companion until it reaches the
Chandrasekhar mass. At this point, thermonuclear burning in the center
of the WD forms a subsonic deflagration flame which---mediated by
thermal conduction of the degenerate electrons---propagates
outward. Since the resulting stratification of dense fuel and
light ashes in the gravitational field is unstable (Rayleigh-Taylor
instability), burning bubbles of hot ashes ascend into cold fuel. At
the interfaces a secondary shear instability gives rise to the local
development of turbulence which wrinkles the flame front. This effect
accelerates the effective burning velocity and thus the energy
generation can account for SN Ia explosions. In the explosion process
the WD material is converted to iron group elements and a smaller
fraction of intermediate-mass elements (like Si, S, and Ca). However,
it is only the radioactive decay of $^{56}$Ni that powers the observed
lightcurve.

Numerical implementations of such models must fulfill a number of
requirements. They have to be robust against variations of the
initial conditions in an astrophysically reasonable range, but on
the other hand they are expected to explain the observed
diversity of SNe Ia. The final goal is, of course, to explain the
correlation between the peak luminosity and the light curve shape on
the basis of theoretical models. This relation is of great importance
to calibrate cosmological distance measurements. Three-dimensional SN
Ia explosion simulations have reached a quality where these issues can
be addressed. Furthermore, nucleosynthetic post-processing of the
explosion data has opened the possibility to calculate synthetic light
curves and spectra which can be directly compared to
observations. This provides a way to discriminate between different
astrophysical models (e.g.\ pure deflagration or delayed detonation). 

We present the first systematic study on what answers
three-dimensional deflagration models can give. What are the possible
parameters that have the potential to explain the SN Ia diversity?
Among others the progenitor's carbon-to-oxygen ratio, its metallicity,
and the central density at ignition are commonly suggested. In our
survey we vary these three parameters independently to explore the
effects on the explosion models. However, we are aware of the fact
that in principle they are interrelated by stellar evolution of the progenitor
WD star. Our study covers the following parameter
space: We apply three different carbon mass fractions of the WD
material, $X(^{12}\mathrm{C}) = 0.30, 0.46, 0.62$, three different
central densities at ignition, $\rho_c = [1.0, 2.6, 4.2] \times 10^9
\, \mathrm{g} \, \mathrm{cm}^{-3}$, and three different metallicities
of the WD (represented by the $^{22}$Ne mass fraction), $Z = [0.3,
1.0, 3.0] Z_\odot$. This defines the 27 models of our survey.

\begin{figure}[t]
  \centerline{\epsfxsize=0.9\textwidth\epsffile{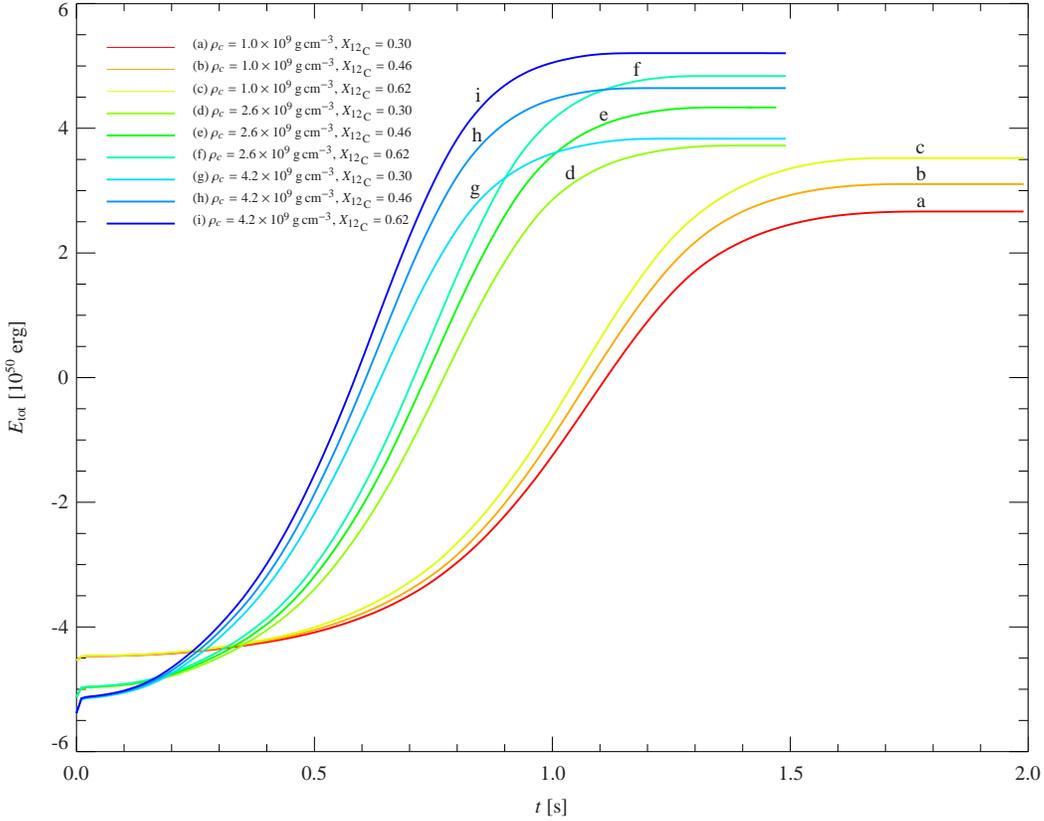}}
  \caption{Total energy of the different models as a function of time.}
  \label{roepke.fig1}
\end{figure}

\subsection{Numerical models}

Our numerical model is based on the \textsc{Prometheus} implementation
\cite{roepke.4}
and describes the flame as a discontinuity between fuel and ashes
applying the level set method \cite{roepke.5}. Turbulence on
unresolved scales is treated with a subgrid-scale model
\cite{roepke.6}. Details of the implementation can be found in
\cite{roepke.7, roepke.8}. The thermonuclear burning is
assumed to proceed to nuclear statistical
equilibrium (NSE) consisting of iron group elements (represented by
``Ni'' in our model) and $\alpha$-particles at high fuel
densities. Below $\rho_{\mathrm{fuel}} = 5.25 \times 10^7 \, \mathrm{g} \,
\mathrm{cm}^{-3}$ burning is assumed to terminate at intermediate mass
elements represented by ``Mg'' and below $\rho_{\mathrm{fuel}} = 1.0 \times 10^7 \,
\mathrm{g} \, \mathrm{cm}^{-3}$ the reactions are so slow that it
is no longer followed. Our numerical setup is analogous to the
\emph{c3\_3d\_256} model of \cite{roepke.8}.
The carbon/oxygen ratio of the progenitor and the central density at
ignition are varied in the explosion models.
In these models tracer
particles are distributed equally in mass shells. They record the
temperature, the density, and the internal energy of the explosion
process. With help of this data it is
possible to post-process the nucleosynthesis of the explosion models
(see C.~Travaglio's contribution to these proceedings). Here the WD's
metallicity is varied by changing the mass fraction
of $^{22}$Ne.

\begin{figure}[ht!]
  \centerline{\epsfxsize= \textwidth\epsffile{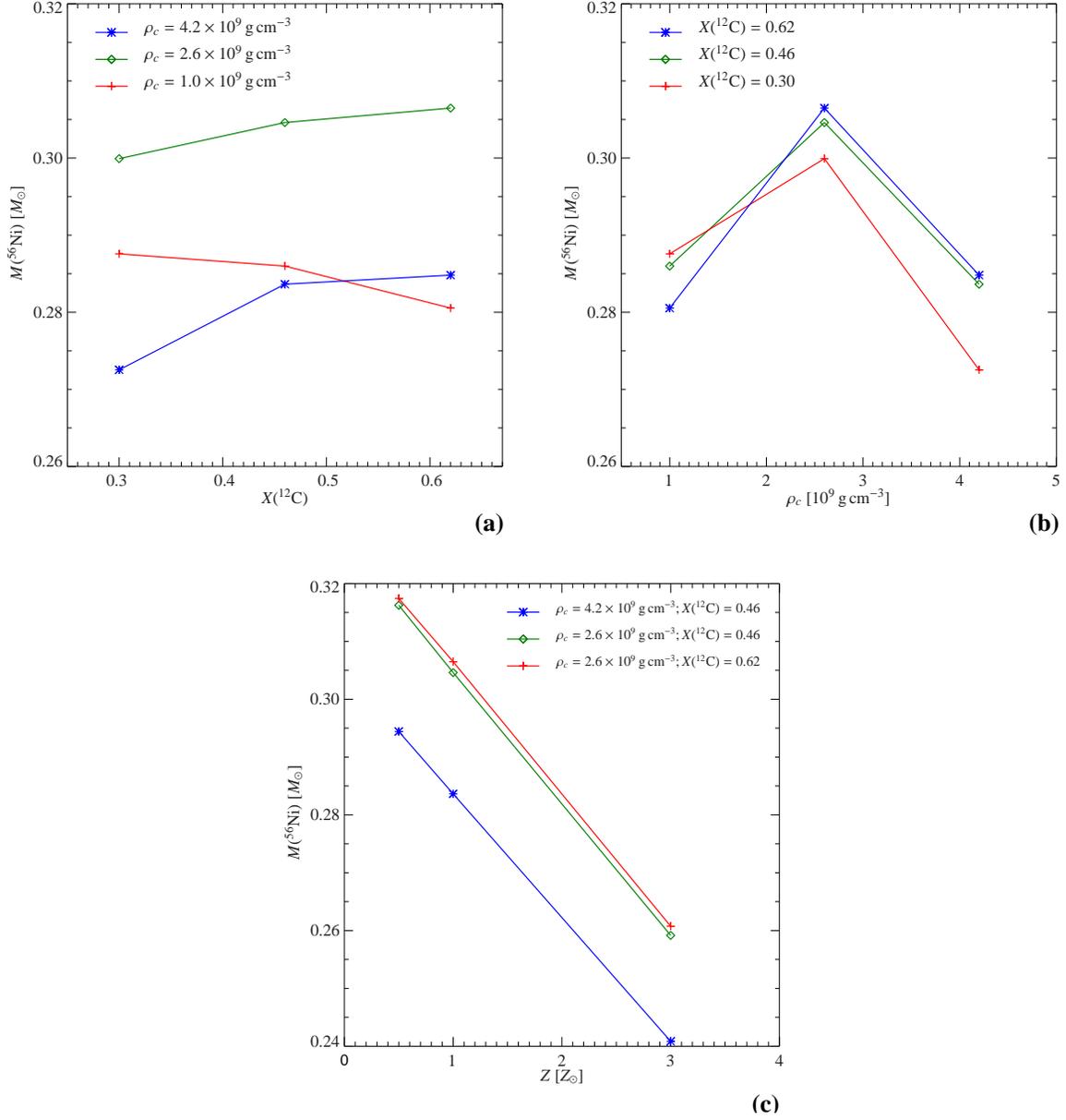}}
  \caption{$^{56}$Ni masses produced by the models as a function
    of (a) carbon mass fraction of the progenitor, (b) central density
    at ignition, and (c) progenitor's metallicity .}
  \label{roepke.fig2}
\end{figure}

\subsection{Results}

Figure \ref{roepke.fig1} shows the total energy of our
models. Obviously, a lower central density leads to a lower energy
release and delays the evolution of the model. The differences in the
total energy productions for varying $\rho_c$ and fixed other
parameters amount to $\sim$40\%. The reason for this effect is the
different gravitational acceleration experienced the flame front which leads to
a change in the evolution of the nonlinear Rayleigh-Taylor
instabilities. For higher $\rho_c$ the evolution is faster and more
pronounced, so that the flame is affected by stronger turbulence. This
accelerates the flame propagation and leads to an increased energy release.

Contrary to that, a change in the
carbon mass fraction does not lead to a significant delay of the
evolution. The change of the total energy production of the models
amounts to $\sim$12\%. This is due to the different binding energies
of fuel with varying C/O ratio.

The $^{56}$Ni masses produced by the different models are plotted in
Fig.~\ref{roepke.fig2}. 
With increasing central density of the WD at ignition, the $^{56}$Ni
production rises to a maximum at $\rho_c \approx 2.5 \times 10^9 \,
\mathrm{g} \, \mathrm{cm}^{-3}$ and then declines again (cf.\
Fig.~\ref{roepke.fig2}b). The variation 
is of the order of 10\%. This behavior can be
expected from the fact that two effects are competing here. The effect
of the different gravitational acceleration experienced by the flame
leads to a faster flame propagation for higher $\rho_c$, as discussed
above. Consequently, more material is processed at high densities
resulting in an increased amount of iron group elements. On the other
hand, at higher densities the
neutronization of the burnt material becomes important and thus an increasing
fraction of the iron group elements is present in form of neutron-rich
nuclei (like $^{58}$Ni) instead of $^{56}$Ni.

With increasing metallicity of the progenitor, i.e.\ higher mass fraction
of $^{22}$Ne (a nucleus with neutron excess), neutron-rich iron group
nuclei in the ashes are favored and less $^{56}$Ni is produced. This is evident in our models (cf.\
Fig.~\ref{roepke.fig2}c). The variation amounts to $\sim$20\%. Our
results are consistent with a linear relation that was proposed by
\cite{roepke.9} (see also the contribution of E.~Brown to these
proceedings).

With varying $X(^{12}\mathrm{C})$ the change
in the nickel masses is only of the order of a few percent (see
Fig.~\ref{roepke.fig2}a).
This result is
unexpected in a simple picture. Although the total explosion
energy increases with $X(^{12}\mathrm{C})$, the flame evolution in the
different models is found to be surprisingly similar. This results in
little changes in the $^{56}$Ni production. The explanation of this
effect is a higher fraction of $\alpha$-particles in NSE in the ashes
at maximum energy generation for increasing $X(^{12}\mathrm{C})$. This
acts as an energy buffer due to the reduction of the binding energy
and increased particle number in the burnt material. Details can be
found in \cite{roepke.10}.

\subsection{Conclusions}

In the present study it has been shown that our SN Ia models are
robust to astrophysically reasonable variations in the initial
parameters. To what degree the deflagration scenario is able to
explain the full observed SN Ia sample has to be decided by more
detailed studies. The current survey indicates that it can account for
some of the observed features.
A simple deduction of the peak luminosities from the $^{56}$Ni masses
(Arnett's rule, \cite{roepke.11}) produced by our models leads a
diversity in the luminosities that can---at least partly---explain
the observed scatter.  

However, it would predict a significant change with varying
metallicity but almost no change with
varying C/O ratio of the progenitor WD. On the other hand, the former
will not influence the explosion dynamics while the latter leads to a
significant change in the total energy release. Thus the light curve
shape will probably not change much in the first case but vary
significantly in the second. Therefore, it is clear that a single
parameter
will not explain the peak luminosity--light curve shape relation of
SNe Ia. It seems likely that a combination of progenitor parameters
(based on stellar evolution models of the WD) will be necessary for
this task. Additionally, synthetic light curves from the models are
needed to to deduce their characteristics and to take into account
multidimensional effects. Arnett's rule
may be a too strong simplification here. One of the next steps in
improving the explosion models will be a refined description of the
thermonuclear reactions. Moreover, it will be necessary to take into
account electron captures at high densities to improve the reliability
of the results for varying central densities at ignition. A detailed
analysis of the parameter study presented here can be found in
\cite{roepke.12}.

\subsection*{Acknowledgements}

The junior authors of this text would like to congratulate
W.~Hillebrandt on his 60th birthday.



\bbib
\bibitem{roepke.1} M.~Reinecke, W.~Hillebrandt, and J.C.~Niemeyer, A\&A
    {\bf 391} (2002) 1167.
\bibitem{roepke.2} V.N.~Gamezo, A.M.~Khokhlov, E.S.~Oran,
  A.Y.~Chtchelkanova, and R.O.~Rosenberg, Science
    {\bf 299} (2003) 77.
\bibitem{roepke.3} W.~Hillebrandt and J.C.~Niemeyer,
  Annu. Rev. Astron. Astrophys.
    {\bf 38} (2000) 191.
\bibitem{roepke.4} B.A.~Fryxell and E.~M{\"u}ller, MPA Green Report
    {\bf 449} (Max-Planck-Institut f{\"ur} Astrophysik, Garching, 1989).
\bibitem{roepke.5} S.~Osher and J.A.~Sethian, J. Comp. Phys.
    {\bf 79} (1988) 12.
\bibitem{roepke.6} J.C.~Niemeyer and W.~Hillebrandt, ApJ
    {\bf 452} (1995) 769.
\bibitem{roepke.7} M.~Reinecke, W.~Hillebrandt, J.C.~Niemeyer,
   R.~Klein, and A.~Gr{\"o}bl, A\&A
    {\bf 347} (1999) 724.
\bibitem{roepke.8} M.~Reinecke, W.~Hillebrandt, and J.C.~Niemeyer, A\&A
    {\bf 386} (2002) 936.
\bibitem{roepke.9} F.X.~Timmes, E.F.~Brown, and J.W.~Truran, ApJ
    {\bf 590} (2002) L83.
\bibitem{roepke.10} F.K.~R{\"o}pke and W.~Hillebrandt, A\&A Lett.
    (2004) in press.
\bibitem{roepke.11} W.D.~Arnett, ApJ
    {\bf 253} (1982) 785.
\bibitem{roepke.12} F.K.~R{\"o}pke, M.~Gieseler, M.~Reinecke,
  C.~Travaglio, and W.~Hillebrandt, (2004) in preparation.
\ebib
\end{document}